\documentclass[notitlepage,showkeys,aps,prd,letterpaper,superscriptaddress]{revtex4-1}

\usepackage{graphicx}

\begin{document}

\sloppy

\title{Search in 8 TeV proton-proton collisions with the MoEDAL \\ monopole-trapping test array}

\author{K. Bendtz}
\affiliation{Department of Physics, Stockholm University, 106 91 Stockholm, Sweden}

\author{A. Katre} 
\affiliation{Particle Physics department, University of Geneva, 1211 Geneva 4, Switzerland}

\author{D. Lacarr\`ere}
\affiliation{Physics Department, CERN, 1211 Geneva 23, Switzerland}

\author{P. Mermod}
\thanks{Corresponding author. E-mail: philippe.mermod@cern.ch}
\affiliation{Particle Physics department, University of Geneva, 1211 Geneva 4, Switzerland}

\author{D. Milstead}
\affiliation{Department of Physics, Stockholm University, 106 91 Stockholm, Sweden}

\author{J. Pinfold}
\author{R. Soluk}
\affiliation{Physics Department, University of Alberta, Edmonton, Alberta, Canada T6G 0V1}

\begin{abstract}
The magnetic monopole appears in theories of spontaneous gauge symmetry breaking and its existence would explain the quantisation of electric charge. MoEDAL is the latest approved LHC experiment, designed to search directly for monopoles produced in high-energy collisions. It has now taken data for the first time. The MoEDAL detectors are based on two complementary techniques: nuclear-track detectors are sensitive to the high-ionisation signature expected from a monopole, and the magnetic monopole trapper (MMT) relies on the stopping and trapping of monopoles inside an aluminium array which is then analysed with a superconducting magnetometer. The first results obtained with the MoEDAL MMT test array deployed in 2012 are presented. This experiment probes monopoles carrying a multiple of the fundamental unit magnetic charge for the first time at the LHC. 
\end{abstract}


\maketitle


\section{Introduction}
\label{intro}

The existence of monopoles would add symmetry to Maxwell's equations of electromagnetism. In 1931, Dirac showed that electric charge quantisation could be explained as a natural consequence of angular momentum quantisation if one assumes the existence of magnetic monopoles~\cite{Dirac1931}. In 1974, 't Hooft and Polyakov independently demonstrated that a GUT scheme with the U(1) subgroup of electromagnetism embedded into a larger gauge group which becomes spontaneously broken by the Higgs mechanism automatically possessed a topological monopole solution~\cite{tHooft1974,Polyakov1974}. It has also been proposed that such a monopole solution could arise in the electroweak theory itself~\cite{Cho1997}. This so-called electroweak monopole would have a mass of the order of several TeV~\cite{Kirkman1981}, possibly within reach of the LHC. From an experimental point of view, these considerations are compelling motivations for searching for free magnetic charges in nature, and the monopole mass is treated as a free parameter. 

It follows from Dirac's argument that monopoles should carry a magnetic charge $g$ equal to a multiple of a fundamental unit of magnetic charge referred to as the Dirac charge $g_D$: $g=n\cdot g_D$, where $g_D$ is equivalent to 68.5 times the charge of an electron. The minimum value of the quantisation number is $n=1$ according to Dirac or $n=2$ according to Schwinger~\cite{Schwinger1975}, and $n=3$ or $n=6$ if one considers the elementary charge to be carried by the down-quark. As a result of the high value of the Dirac magnetic charge, a high-velocity monopole is expected to suffer energy losses in matter over $4500$ times higher than a muon~\cite{Ahlen1978,Ahlen1980,Ahlen1982}. Monopoles should thus generally manifest themselves as very highly ionising particles. Another way to identify a free magnetic charge is to measure the persistent current it would induce when passed through a superconducting loop. Searches for monopoles using these two signatures (high ionisation and magnetic induction) were performed in cosmic rays and in matter, and at accelerators each time a new energy regime was reached~\cite{Fairbairn2007}. No evidence for their existence has been gathered so far. 

At colliders, three kinds of techniques are commonly used: (1) General-purpose detectors with high ionisation energy loss detection capabilities (e.g. OPAL at LEP~\cite{OPAL2008} and CDF at the Tevatron~\cite{CDF2006}); (2) dedicated deployment of nuclear-track detectors~\cite{Pinfold2009} around the interaction points (e.g. at LEP~\cite{Kinoshita1992,Pinfold1993} and at the Tevatron~\cite{Bertani1990}); and (3) the induction technique applied to accelerator and detector material in which monopoles would have stopped and remained trapped (e.g. at HERA~\cite{H12005} and at the Tevatron~\cite{Kalbfleisch2000,Kalbfleisch2004}). Together, these searches excluded the presence of monopoles with charge equal to or above the Dirac charge and masses up to $400$ GeV. Masses higher by one order of magnitude (up to 4 TeV) can be probed at the LHC. For optimum results, the LHC programme should include all three of these techniques~\cite{DeRoeck2012a}. An initial monopole search was performed at the ATLAS general-purpose experiment~\cite{ATLAS2012a}. Monopole trapping experiments were shown to be feasible at the LHC~\cite{DeRoeck2012b}. The dedicated MoEDAL experiment near the LHCb interaction point uses a combination of in-flight detection with nuclear-track detectors~\cite{MoEDAL2009} and trapping with aluminium absorbers~\cite{Mermod2013}. In this paper, we present the first data from the MoEDAL magnetic monopole trapper (MMT) test array deployed in 2012.

\section{Experimental procedure}
\label{experiment}

The MoEDAL MMT is an aluminium volume placed in the vicinity of the LHCb interaction point. The choice of material is driven by several factors: aluminium is cheap, non-magnetic, and has a nucleus which does not activate and which, thanks to its large spin, would be expected to strongly bind with monopoles which would range out and stop within the array~\cite{Milton2006}. An initial MMT test array comprising 11 boxes each containing 18 rods of 60 cm length and 2.54 cm diameter was installed in September 2012 and exposed to 0.75 fb$^{-1}$ of 8 TeV proton-proton collisions. The boxes were stacked behind the LHCb VELO vacuum vessel just under the beam pipe $\sim$1.8 m away from the interaction point, covering 1.3\% of the total solid angle. After the run was finished, the rods were retrieved and cut into 20 cm samples (except for Box 11, whose rods were cut into a mix of 10, 15, 20 and 30 cm samples) for analysis with a superconducting magnetometer. 

A DC-SQUID rock magnetometer housed at the Laboratory for Natural Magnetism (ETH Zurich) was used. Previous studies performed with rocks and with a small set of material samples from the LHC accelerator demonstrated that this instrument has the capability to detect monopoles trapped in matter with charges much less and much larger than the Dirac charge~\cite{DeRoeck2012b,Bendtz2013a}. The calibration was performed with a convolution method applied to a dipole sample, and cross-checked using thin solenoids which mimic a monopole of well-known magnetic charge~\cite{DeRoeck2012b}. The magnetometer response is linear and charge symmetric. This allows one to express the measured currents in units of Dirac magnetic charge. 

Each of the 606 samples was passed at least once through the magnetometer, mostly during a measurement campaign in September 2013. Every tenth measurement on average was performed with an empty sample holder for the offset subtraction. The relevant measured value is the persistent current, defined as the difference between the currents measured after and before passage of the sample through the sensing coil, and then subtracting the difference obtained with a nearby empty holder measurement. Whenever the persistent current differs from zero by more than one fourth of a Dirac charge --- spurious offset jumps can cause this to happen in $\sim$2\% of the measurements --- the sample is considered a candidate and measured again several times.  A sample containing a genuine monopole would consistently yield the same value for repeated measurements, while values consistent with zero are expected whenever an instrumental effect occurred in the first measurement. Including first and multiple sample measurements as well as empty holder and calibration measurements, a total of 852 independent measurements were performed in 7 days.

\section{Results of magnetometer measurements}
\label{results}

\begin{figure}[tb]
\begin{center}
  \includegraphics[width=0.95\linewidth]{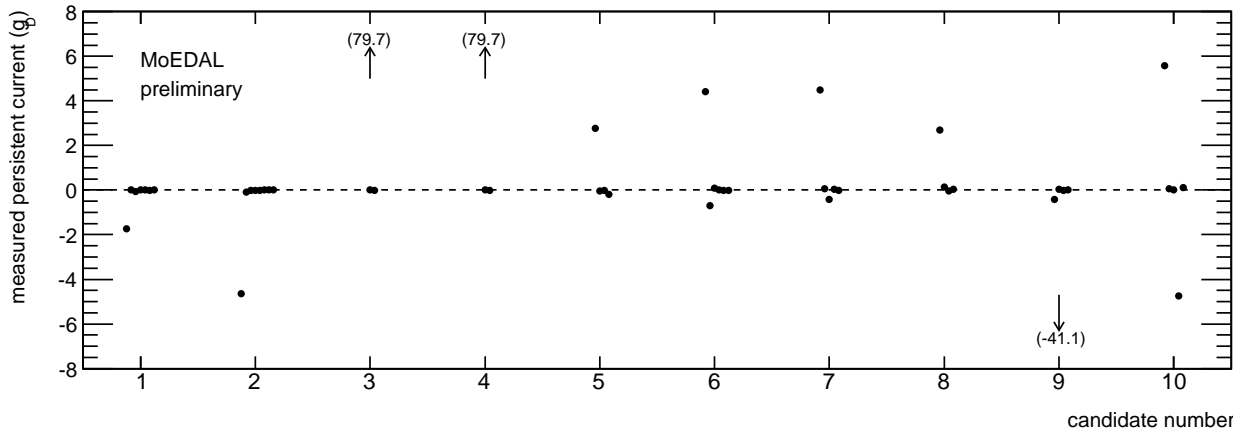}
  \caption{Results of multiple persistent current measurements (in units of the Dirac charge) for the 10 samples which yielded large ($>0.25~g_D$) values for the first measurement.}
\label{fig:MMcandidates}
\end{center}
\end{figure}

Results for potential candidates measured multiple times are shown in Fig.~\ref{fig:MMcandidates}. In each case where the first measurement showed a large jump, additional analysis of the same sample was consistent with a zero magnetic charge. It was also noticed that the jumps occurred more often for certain periods during which the magnetometer response was less stable than usual. Such instabilities could be caused by a number of instrumental and environmental factors: trapped magnetic flux in the SQUID, small ($\sim$mm) variations in the length of the sample holder from one run to another, the accumulation of condensed water and ice in the magnetometer tube near the cold sensing region, physical vibrations and shocks, imperfect shielding from magnetic fields produced by external electrical devices, and induced currents in the samples when they move quickly through the sensing coils. With experience, measures can be taken to minimise such effects when performing measurements. 

\begin{figure}[tb]
\begin{center}
  \includegraphics[width=0.65\linewidth]{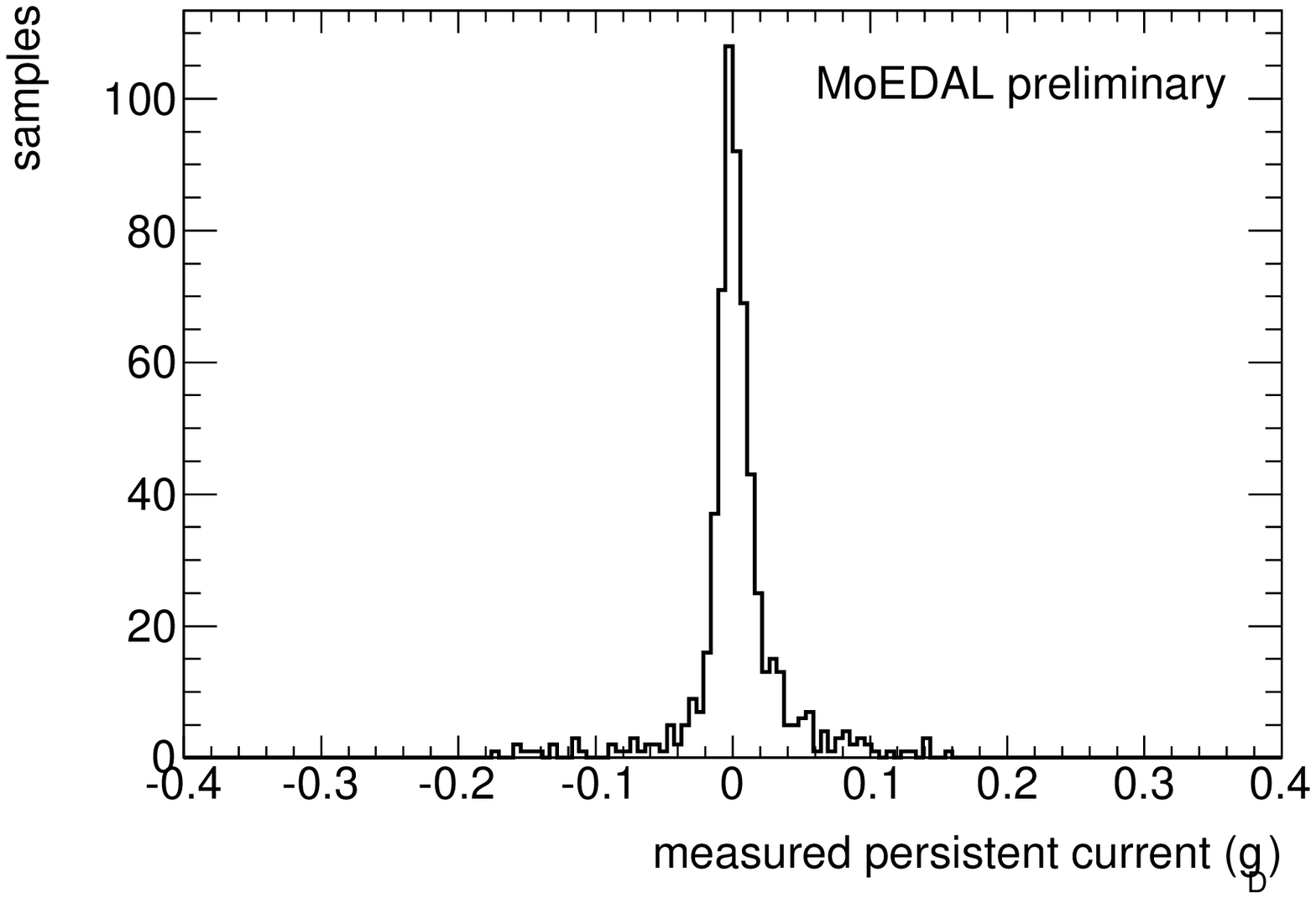}
  \includegraphics[width=0.99\linewidth]{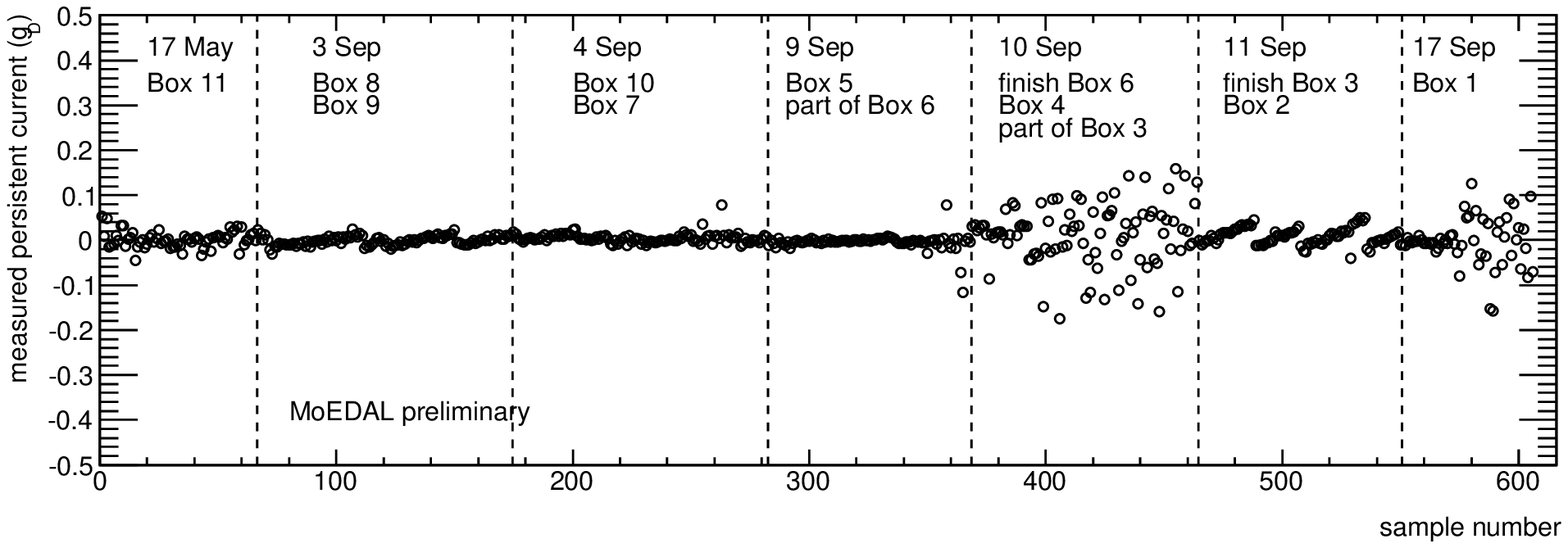}
  \caption{Magnetic charge (in units of the Dirac charge) measured in the 606 aluminium samples of the 2012 MMT test array. }
\label{fig:MMTtest}
\end{center}
\end{figure}

The magnetic charge contained in all 606 samples of the MMT test array --- as measured by the first measurement or a subsequent measurement in the cases where a spurious offset jump was observed for the first measurement --- is shown in Fig.~\ref{fig:MMTtest}. The bottom plot gives an idea of the evolution of the resolution with time, where periods of relative instability are observed for the 10th and 17th of September. The small escalator feature (most visible for 11th of September) is due to offset drifts~\cite{DeRoeck2012b} being corrected for sporadically using empty holder measurements. The top plot shows the same data as a single histogram. The resolution is good enough to exclude the presence of monopoles with charge larger than 0.4 $g_D$ in the samples.

\section{Interpretation of the results}
\label{interpretation}

The amount of material between the interaction point and the MoEDAL arrays is typically less than that in front of EM calorimeters in general-purpose detectors, providing sensitivity to higher magnetic charges and lower energies~\cite{MoEDAL2009,DeRoeck2012a}. Using an average\footnote{More work needs to be done to precisely quantify the material distribution in the cavern and estimate uncertainties due to the approximations made, which will allow one to quote cross-section limits.} of 14 g$\cdot$cm$^{-2}$ of steel for the material of the VELO chamber in front of the MMT test array and 23 g$\cdot$cm$^{-2}$ of aluminium for the array itself, calculations of monopole stopping in the material~\cite{DeRoeck2012a} show that the array would trap more than 0.5\% of monopoles with charge between $n=1$ and $n=4$ and mass up to 3500 GeV when assuming a kinematic distribution consistent with a Drell-Yan pair production model. By comparison, the ATLAS published results for 7 TeV collisions~\cite{ATLAS2012a} --- until now the only existing constraints on monopoles at the LHC --- are quoted only for $n=1$ and a mass up to 1500 GeV.

\section{Conclusions and outlook}
\label{conclusions}

MoEDAL is designed for passive detection of magnetic monopoles, both in-flight (with the track-etch technique) and trapped (with the induction technique). A magnetic monopole trapping (MMT) test array was exposed to 0.75 fb$^{-1}$ of 8 TeV proton-proton collisions in 2012. Full scanning of this array with a superconducting magnetometer was performed and no monopoles were found in any of the samples. 

Despite a small solid angle coverage and modest luminosity, the MoEDAL MMT subdetector provides an opportunity to probe ranges of charge, mass and energy which are not currently accessible to other experiments. Furthermore, this technique can yield results very quickly and would allow for a robust, background-free assessment of a signal, potentially providing a direct measurement of a monopole magnetic charge based on its electromagnetic properties only. The full-scale MMT array will be more than 4 times larger than the test array and is scheduled for deployment in $2014-2015$ along the back and rear of the LHCb VELO vessel as well as under the floor below the interaction point. This experiment is expected produce the first monopole search results in 14 TeV collisions.

\section*{Acknowledgements}

This work was supported by a fellowship from the Swiss National Science Foundation and a grant from the Marc Birkigt Fund of the Geneva Academic Society.

\bibliographystyle{mystylem}
\bibliography{ComoMermod2013}

\end{document}